\documentclass{article}

\usepackage[preprint,nonatbib]{neurips_2022}

\usepackage{graphicx}
\usepackage[utf8]{inputenc} 
\usepackage[T1]{fontenc}    
\usepackage[symbol]{footmisc}
\usepackage{hyperref}       
\usepackage{url}            
\usepackage{booktabs}       
\usepackage{amsfonts}       
\usepackage{nicefrac}       
\usepackage{multirow}
\usepackage{tabularx}
\usepackage{caption}
\usepackage{amssymb}
\usepackage{amsmath}
\usepackage{bbold}                              
\usepackage{microtype}
\usepackage{siunitx}
\usepackage{etoolbox}                           
\newcommand{\ubold}{\fontseries{b}\selectfont}  
\robustify\ubold                                

\newcolumntype{C}{>{\centering\arraybackslash}X}
\newcolumntype{L}{>{\raggedright\arraybackslash}X}
\newcommand{\tablecaptionsep}{\vspace*{1pt}}

\usepackage{xcolor}
\definecolor{blueemph}{HTML}{000000}
\newcommand{\blueemph}[1]{\textcolor{blueemph}{#1}}
\definecolor{greenemph}{HTML}{000000}
\newcommand{\greenemph}[1]{\textcolor{greenemph}{#1}}

\definecolor{hermancolor}{HTML}{FF6600}

\definecolor{mattcolor}{HTML}{006e05}

\title{TransFusion: \\ Transcribing Speech with Multinomial Diffusion}

\renewcommand{\thefootnote}{\fnsymbol{footnote}}

\author{%
  Matthew Baas \protect\footnote[2]{Equal contribution} \\
  \texttt{20786379@sun.ac.za} \\
   \And
   Kevin Eloff \protect\footnotemark[2] \\
   \texttt{20801769@sun.ac.za} \\
   \And
   Herman Kamper \\
   \texttt{kamperh@sun.ac.za}\\
   \AND

  \textrm{\normalfont MediaLab, Department of Electronic \& Electrical Engineering}\\
  Stellenbosch University\\
  South Africa \\
}
\def\modelname{{TransFusion}}

\begin{document}

\maketitle

\footnotetext[2]{Equal contribution.}
\renewcommand*{\thefootnote}{\arabic{footnote}}

\begin{abstract}
Diffusion models have shown exceptional scaling properties in the image synthesis domain, and initial attempts have shown similar benefits for applying diffusion to unconditional text synthesis.
Denoising diffusion models attempt to iteratively refine a sampled noise signal until it resembles a coherent signal (such as an image or written sentence).
In this work we aim to see whether the benefits of diffusion models can also be realized for speech recognition.
To this end, we propose a new way to perform speech recognition using a diffusion model conditioned on pretrained speech features.
Specifically, we propose TransFusion:
a transcribing diffusion model which iteratively denoises a random character sequence into coherent text corresponding to the transcript of a conditioning utterance.
We demonstrate comparable performance to existing high-performing contrastive models on the LibriSpeech speech recognition benchmark.
To the best of our knowledge, we are the first to apply denoising diffusion to speech recognition.
We also propose new techniques for effectively sampling and decoding multinomial diffusion models.
These are required because traditional methods of sampling from acoustic models are not possible with our new discrete diffusion approach. Code and trained models are available: \url{https://github.com/RF5/transfusion-asr}.
\end{abstract}

\section{Introduction}\label{sec:1_intro}

Automatic speech recognition (ASR) is the task of 
transcribing a speech utterance into the words being said.
The current paradigm for high-performance ASR involves the use of supervised training of large neural networks with a connectionist temporal classification (CTC) loss \cite{ctc}.
Intuitively,
these models
predict a probability of a character occurring in a particular time window within an utterance.
While this method is the current state-of-the-art for ASR \cite{chen2022wavlm, chung2021w2v-bert}, the question remains whether better methods might exist.
We aim to approach ASR from a new perspective and evaluate how closely such an initial attempt can approach the state-of-the-art CTC-based models.

We have seen in other domains such as image and audio synthesis \cite{imagen_saharia2022photorealistic, diffwave_kong2020} that denoising diffusion probabilistic models (or `diffusion models') \cite{diffusion_sohl2015deep} have exceptional scaling and performance properties. 
Diffusion models are trained to iteratively denoise a signal sampled from a known noise distribution until it resembles a coherent signal of interest (e.g. images or audio).
Efforts like \cite{clas_free_guidance_ho2021, dall-e-2_ramesh2022hierarchical} have shown that we can condition this denoising process to correspond to some desired signal  (e.g. a text description of an image).
In this work we aim to determine whether applying diffusion to speech recognition yields similar properties, and to what extent it can compete with the current \mbox{best CTC-type models.}

Concretely, we attempt to formulate ASR as a conditional diffusion task.
Conditioned on speech features from a self-supervised speech representation model, our system attempts to iteratively denoise a random character sequence to ultimately resemble the transcript of the utterance associated with the speech features.
Self-supervised speech representation models process speech into a sequence of vectors that represent high-level information about the speech \cite{chen2022wavlm}.
Our model uses these speech features as conditioning in a multinomial diffusion task~\cite{multinomial_diff_NEURIPS2021_67d96d45} -- a discrete variant of diffusion -- whereby the model predicts a distribution of characters occurring at each position in a transcript.
Since our model transcribes speech with a diffusion task, we dub it \modelname{}.
To the best of our knowledge, we are the first to apply diffusion to the task of ASR.
Furthermore, the typical decoding methods used in ASR are not readily applicable to our new type of model.
So, we also go on to propose initial new techniques for improving sampling of diffusion-type ASR acoustic models.

We compare our model to existing high-performing CTC-type models such as wav2vec 2.0 \cite{wav2vec2.0} on the standard LibriSpeech ASR benchmark \cite{panayotov2015librispeech}.
We do not use a language model or external lexicon, as typical methods of combining acoustic and language models have not yet been developed for our new diffusion approach.
In this evaluation setting, we
demonstrate comparable word error rate (WER) performance to existing high-performing CTC-type models of similar size (\texttt{test-other} WER of 8.8\%) despite the dearth of decoding and sampling heuristics available for our new diffusion-type ASR approach.
In summary, we find that the scaling properties of diffusion models in other domains are also present in ASR.
We also recognize the need for future development of larger ASR diffusion models and for
methods to combine language models with diffusion acoustic models. 
Code, models, and demo: {\footnotesize \url{https://github.com/RF5/transfusion-asr}}.
\section{Related Work}\label{sec:related}

Modern high-performance ASR systems operate in the time-domain, typically using end-to-end deep neural networks to transcribe an utterance.
In particular, current state-of-the-art methods such as \cite{chung2021w2v-bert, hubert2021, wav2vec2.0} first use a large convolutional encoder to downsample a waveform into a vector sequence with each vector typically corresponding to 10~ms to 50~ms of audio.
This sequence is then refined in a large transformer variant \cite{attention_is_all_you_need} to yield output features.
These models are also typically trained in two phases: a pretraining phase with unlabeled speech, and a fine-tuning phase with labeled speech (i.e. audio where the transcript is known). 
The pretraining phase is formulated in a variety of ways, but often involves a contrastive or masked language modelling task whereby these output features must accurately predict what information is present in that time window even if that portion of the input audio is masked \cite{wav2vec2.0, hubert2021}.
The fine-tuning process to perform ASR is \mbox{done with CTC, discussed next.}

Meanwhile, diffusion models (outlined later) have been almost exclusively applied to continuous domains such as image synthesis \cite{imagen_saharia2022photorealistic, dall-e-2_ramesh2022hierarchical} and music or audio synthesis \cite{diffwave_kong2020}. 
Movellan et al. (1999)~\cite{movellan1999diffusion_old} was the first to apply diffusion with textual data, where they attempted to classify the word spoken in short videos of people saying one of four possible words.
However, the diffusion framework referenced in this work is \textit{not} a denoising diffusion probabilistic model.
Rather, they define their own concept of `diffusion networks' as a continuous stochastic version of recurrent neural networks \cite{movellan1999diffusion_old}.
The second, more recent, work considering diffusion with textual data defined the key formulation for diffusion on discrete units such as text characters -- aptly named multinomial diffusion \cite{multinomial_diff_NEURIPS2021_67d96d45}.
While the authors of \cite{multinomial_diff_NEURIPS2021_67d96d45} show considerable performance at unconditional text synthesis, they leave the question open as to how effective such diffusion methods will be when applied to ASR~--~the goal of this work.

\subsection{Connectionist Temporal Classification}
The fine-tuning step of most existing end-to-end ASR systems involves the use of connectionist temporal classification (CTC).
CTC is a method to model the probability of one sequence given a different, possibly unaligned sequence \cite{ctc}.
For speech recognition, the sequence of output features produced by the pretraining step discussed earlier is used to model the probability of the sequence of characters (target transcript).
These sequences are unaligned since each output feature from the model corresponds to a small window of time (e.g. 10~ms), while an English character may correspond to a long period of time (e.g. 300~ms).
Essentially, each item in the output sequence produced by the model parameterizes a categorical distribution over the characters in the alphabet and a special $\epsilon$ character. 
CTC then allows for many-to-one alignments of this output sequence to the ground-truth transcript by collapsing repeated characters and removing $\epsilon$ characters.
A loss is formed as the negative log likelihood of all possible alignments between the model output and target character sequence, where dynamic programming is used to make the computation tractable \cite{ctc_explained_hannun2017sequence}.
So, the fine-tuning process of the current large self-supervised models such as \cite{wav2vec2.0, hubert2021} involves maximizing the likelihood of the ground-truth transcript, given the model's output features. 

\subsection{Denoising Diffusion Probabilistic Models}

One of the newer techniques that is rising in popularity for speech and image synthesis models is that of denoising diffusion probabilistic models, or simply `diffusion models'. 
Concretely, a diffusion model \cite{diffusion_sohl2015deep} defines a Markov process of $T$ steps from $t \in \{0,...,T - 1\}$. 
The modelled data (e.g. waveforms or images) is defined
as
the signal at the first timestep $\mathbf{x}_0$, and the last timestep is defined as a known noise distribution, e.g. $\mathbf{x}_T = \mathcal{N} (\mathbf{0}, \mathbf{I})$ for images. 
The diffusion process consists of a \textit{forward} and \textit{reverse} function to move the signal from $\mathbf{x}_t$ to $\mathbf{x}_{t+1}$ and from $\mathbf{x}_t$ to $\mathbf{x}_{t-1}$, respectively. 
The forward diffusion process is defined by a function $q(\mathbf{x}_t | \mathbf{x}_{t-1})$ which iteratively \textit{adds} noise to a signal until -- at timestep $T$ -- it resembles a pure noise distribution. 
Similarly, the reverse diffusion process $p(\mathbf{x}_{t-1}|\mathbf{x}_t)$ iteratively denoises the signal until at $t=0$ it resembles a coherent signal. 
This reverse process is parameterized with a large neural network \cite{diffusion_sohl2015deep}.

Specifically, the diffusion network (the model associated with the reverse diffusion process $p$) is trained to predict the noise added through the forward process -- i.e. to predict the difference between the desired coherent signal and the signal after adding varying amounts of noise. 
At each inference step the diffusion network is called to parameterize $p(\mathbf{x}_{t-1}|\mathbf{x}_t)$ and then sample the slightly denoised next step $\mathbf{x}_{t-1}$.
Diffusion models have recently been shown to scale very well to large model sizes and datasets \cite{imagen_saharia2022photorealistic, dall-e-2_ramesh2022hierarchical}, and we hypothesize that it will yield similar beneficial properties when applied to ASR.
One issue with typical diffusion is that it is formulated in the continuous domain, such as denoising a continuous pixel or audio sample value slightly in each step.
For discrete signals like text, we must use a recent discrete variant of diffusion -- multinomial diffusion.

\subsection{Multinomial Diffusion}\label{sec:multinomial_diff}

Diffusion is typically used for continuous signals such as images or waveforms \cite{imagen_saharia2022photorealistic, diffwave_kong2020}.
However, in 2021, Hoogeboom et al.
introduced a method to perform diffusion on signals with discrete alphabets: multinomial diffusion \cite{multinomial_diff_NEURIPS2021_67d96d45}.

Concretely, multinomial diffusion defines the input to a diffusion model as a sequence of discrete units (i.e. letters or words) represented as one-hot encoded vectors $\mathbf{x}$.
We index the diffusion timestep with $t \in \{0, ..., T-1\}$ and the position in the sequence (the character index in the transcript) with $i \in \{0, ..., N-1\}$ such that
$\mathbf{x}_{t,i} \in \{0, 1\}^K$
is the one-hot encoding of the character represented at diffusion timestep $t$ and sequence position $i$ using a $K$-sized alphabet.
However, all diffusion operations proposed by \cite{multinomial_diff_NEURIPS2021_67d96d45} are independent across sequence length, thus when we omit the index $i$ it indicates that the statement applies to the entire sequence independent of sequence index.
Multinomial diffusion defines the forward noising process $q(\mathbf{x}_{t} | \mathbf{x}_{t-1} )$, the posterior $q(\mathbf{x}_{t-1} | \mathbf{x}_{t}, \mathbf{x}_{0} )$, and the reverse diffusion process $p(\mathbf{x}_{t-1}|\mathbf{x}_{t})$ for a $K$-sized alphabet as:
\begin{equation*}
\begin{split}
 q(\mathbf{x}_{t} \, | \, \mathbf{x}_{t-1} ) &= \mathcal{C} \left(\mathbf{x}_{t} \, \Big| \, (1-\beta_t)\mathbf{x}_{t-1} + \beta_t/K\right) \\ 
 q(\mathbf{x}_{t-1} \, | \, \mathbf{x}_{t}, \mathbf{x}_{0} ) &= \mathcal{C}\left(\mathbf{x}_{t-1} \, \Big| \ \frac{1}{A} \left[\alpha_t \mathbf{x}_{t} + (1 - \alpha_t)/K \right] \odot \left[\bar{\alpha}_{t-1} \greenemph{\mathbf{x}_{0}} + (1 - \bar{\alpha}_{t-1})/K \right] \right) \\
 p(\mathbf{x}_{t-1} \, | \, \mathbf{x}_{t}) &= \mathcal{C}\left(\mathbf{x}_{t-1} \, \Big| \ \frac{1}{A} \left[\alpha_t \mathbf{x}_{t} + (1 - \alpha_t)/K \right] \odot \left[\bar{\alpha}_{t-1} \blueemph{\hat{\mathbf{x}}_{0}} + (1 - \bar{\alpha}_{t-1})/K \right]\right)
\end{split}
\end{equation*}

Where
$\mathcal{C}$ denotes a categorical distribution with category probabilities specified after $|$.
$\beta_t$ is the diffusion noise schedule defined in the original binomial diffusion work \cite{diffusion_sohl2015deep}, and $\alpha_t = 1 - \beta_t$, $\bar{\alpha}_t = \prod_{\tau = 0}^t \alpha_\tau$.
The fraction $\frac{1}{A}$ is a normalizing constant to ensure the probabilities sum to one \cite{multinomial_diff_NEURIPS2021_67d96d45}.
The final-timestep sequence $\greenemph{\mathbf{x}_{0}}$ is the one-hot encoding derived from the ground-truth text for the posterior, and $\blueemph{\hat{\mathbf{x}}_{0}}$ is the predicted probabilities over the vocabulary for each position in the sequence.
This is how the diffusion process is parameterized with a neural network: at each diffusion timestep, the model predicts a distribution over the vocabulary $\hat{\mathbf{x}}_0$ of the \textit{fully denoised transcript} at $t=0$.
To get an intuitive idea how the reverse process iteratively denoises a sample transcript using multinomial diffusion, see Figure~\ref{fig:denoise_example} (explained later).
In this work we phrase the task of speech recognition
as a
speech-feature-guided
multinomial diffusion task.

\section{Model}\label{sec:3_model}

Our model is a denoising probabilistic diffusion model \cite{diffusion_sohl2015deep} that transcribes utterances from a provided sequence of speech features extracted from a self-supervised speech representation model. 
So, as our model is a \underline{trans}cribing dif\underline{fusion} model, we dub it \textbf{\modelname{}}.
Concretely, it
adapts classifier-free guidance \cite{clas_free_guidance_ho2021} and multinomial diffusion \cite{multinomial_diff_NEURIPS2021_67d96d45} to allow a discrete diffusion model to be conditioned on a sequence of speech features extracted from the self-supervised \mbox{speech representation model WavLM~\cite{chen2022wavlm}.}

\begin{figure}[t]
    \centering
    \includegraphics[width=1.0\textwidth]{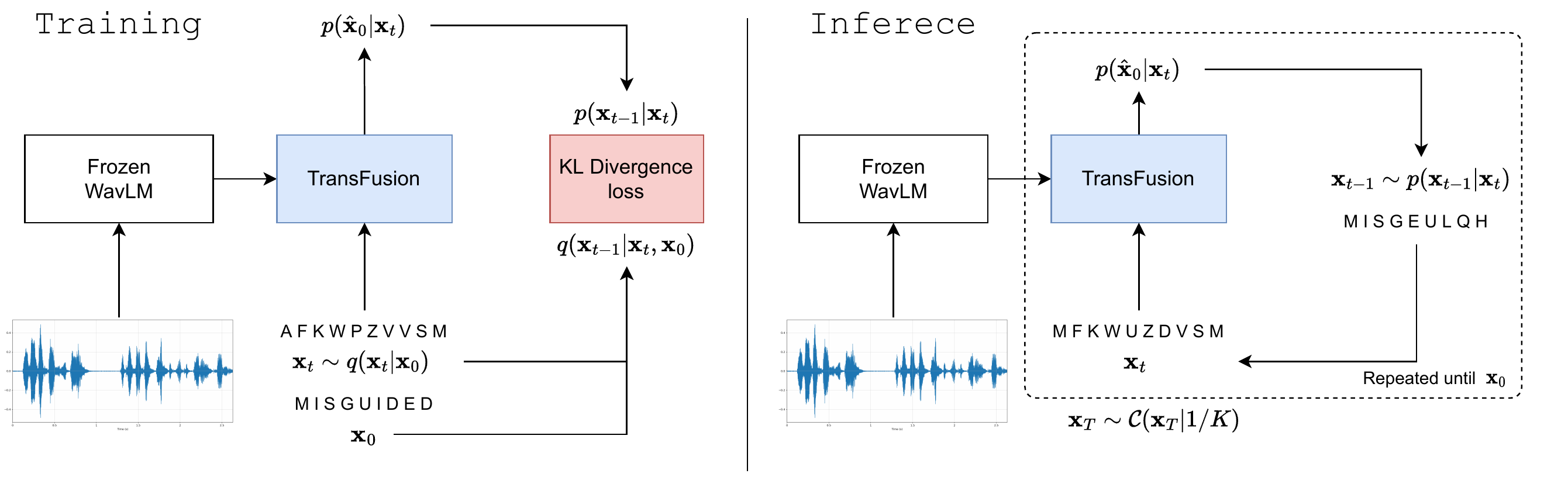}
    \caption{\modelname{} diagram. 
    Speech features from an utterance computed using a frozen WavLM encoder \cite{chen2022wavlm} are used to condition \modelname{}.
    During training (left) the model is trained according to multinomial diffusion \cite{multinomial_diff_NEURIPS2021_67d96d45} to minimize the KL divergence between the reverse process $p(\mathbf{x}_{t-1} | \mathbf{x}_{t})$ and posterior process derived from the ground truth utterance $q(\mathbf{x}_{t-1} | \mathbf{x}_{t}, \mathbf{x}_{0} )$.
    During inference (right), a uniformly random sampled transcript $\mathbf{x}_T$ is iteratively denoised using \modelname{} until $\mathbf{x}_0$ is the predicted output transcript.}
    \label{fig:model_diagram}
    \vspace{-0.2cm}
\end{figure}

\subsection{Conditioning Diffusion on Speech Representations}\label{sec:3_1_conditioning}
The training and inference setup of \modelname{} is shown in Figure~\ref{fig:model_diagram}.
During training, we have an input utterance waveform and its associated ground-truth transcript denoted $\mathbf{x}_0$. 
During each training step, a noised version of the transcript is calculated for diffusion timestep $t$ using $q(\mathbf{x}_t | \mathbf{x}_0) = \mathcal{C}\left(\mathbf{x}_t \, | \, \bar{\alpha}_t \mathbf{x}_0 + (1-\bar{\alpha}_t)/K \right)$ \cite{multinomial_diff_NEURIPS2021_67d96d45}, where $t$ is sampled uniformly at random from $\{0, ..., T-1\}$.
Intuitively, the input text fed to the model has its characters randomly flipped, with increasing randomness until at the highest timestep $t=T-1$, the transcript fed to the model is entirely random.
The waveform is converted into a sequence of high-level speech features $\mathbf{c}$ using a fixed pretrained WavLM model \cite{chen2022wavlm}.
This sequence of features $\mathbf{c}$ is then used to condition the main \modelname{} model.
\modelname{}'s architecture is that of a transformer variant (discussed later) and maps the noisy input characters to a predicted distribution of output characters for the desired transcript $\mathbf{x}_0$.
More formally,

\vspace{-0.1cm}
\begin{equation*}
    p(\hat{\blueemph{\mathbf{x}}}_{0}| \mathbf{x}_t, \mathbf{c}) =  \blueemph{\text{\modelname{}}}(\mathbf{x}_t, \mathbf{c}) 
\end{equation*}

During inference (Figure~\ref{fig:model_diagram}, right), given speech features $\mathbf{c}$ from an utterance with unknown transcript, we sample a random sequence of characters at $\mathbf{x}_T$.
We then iteratively denoise the transcript by using \modelname{} and the diffusion parameters to compute a distribution for the reverse process $p(\mathbf{x}_{t-1} | \mathbf{x}_{t})$.
The slightly denoised transcript is then sampled from this distribution as $\mathbf{x}_{t-1}$ and used as the input to the model in the next iteration.
This process continues until $t=0$ and $\mathbf{x}_0$ is a refined prediction of the transcript of the utterance. 

\subsection{Training Task}

The loss follows that of multinomial diffusion (Section~\ref{sec:multinomial_diff}).
Specifically, using the diffusion parameters $\beta$, $\alpha$, $\bar{\alpha}$ at timestep $t$ and the current noised inputs, the posterior $q(\mathbf{x}_{t-1} | \mathbf{x}_{t}, \mathbf{x}_{0} )$ is computed.
Likewise, with the model's prediction and the diffusion parameters, the reverse distribution $p(\mathbf{x}_{t-1} | \mathbf{x}_{t})$ also provides a distribution over $\mathbf{x}_{t-1}$.
For \modelname{} to accurately undo the noise added between timestep $t-1$ and $t$, the predicted reverse distribution $p(\mathbf{x}_{t-1} | \mathbf{x}_{t})$ should be close to the posterior $q(\mathbf{x}_{t-1} | \mathbf{x}_{t}, \mathbf{x}_{0} )$ (which has access to the ground-truth transcript).
So, a training loss is formed as the Kullback–Leibler (KL) divergence:

\vspace{-0.1cm}
\begin{equation*}
    \mathcal{L} = \text{KL}\left( \, q(\mathbf{x}_{t-1} | \mathbf{x}_{t}, \mathbf{x}_{0} ) \ \| \ p(\mathbf{x}_{t-1} | \mathbf{x}_{t}) \, \right)
\end{equation*}

Furthermore, the theory of multinomial diffusion also requires an additional term be added when $t=0$ \cite{multinomial_diff_NEURIPS2021_67d96d45}.
Namely the cross-entropy between the one-hot ground-truth distribution $\mathbf{x}_0$ and the predicted probabilities from \modelname{} $\hat{\mathbf{x}}_0$ is added to the loss when $t=0$ \cite{multinomial_diff_NEURIPS2021_67d96d45}.
Intuitively, this pushes the distribution predicted by the model to be close to the one-hot ground truth targets when $t=0$.
This loss is readily computed and the model can be trained through backpropagating through the predictions $\blueemph{\hat{\mathbf{x}}_0}$ used to compute $p(\mathbf{x}_{t-1} | \mathbf{x}_{t})$.

\subsection{Architecture}

\begin{figure}[t]
    \centering
    \includegraphics[trim={1.2cm 0 2.5cm 0},clip,width=1.0\textwidth]{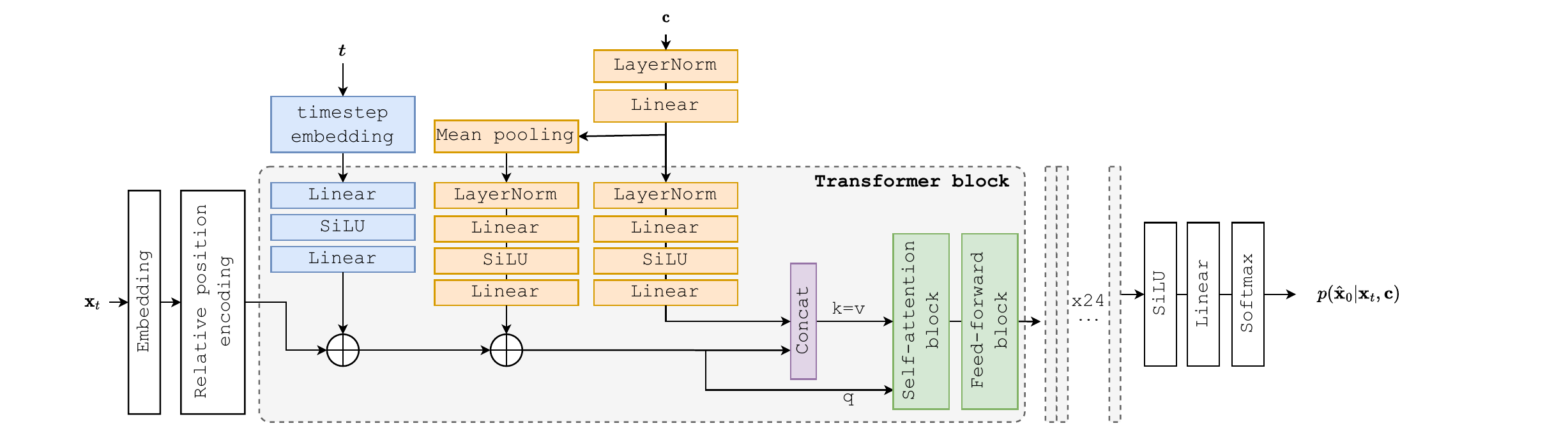}
    \caption{\modelname{} architecture.
    The sequence of input characters $\mathbf{x}_t$ is passed through an embedding and positional encoding layer and then into a sequence of 24 transformer blocks before being projected to an output distribution $p(\hat{\mathbf{x}}_0 | \mathbf{x}_t, \mathbf{c})$ over the vocabulary for each character in the sequence.
    In each transformer block, the mean-pooled WavLM features $\mathbf{c}$ and processed timestep embedding is added to each vector in the sequence which acts as the query to a self-attention block~\cite{attention_is_all_you_need}.
    The key and value are formed by concatenating the transformed sequence of WavLM vectors with the main sequence derived from the characters.
    \texttt{SiLU}, \texttt{LayerNorm}, \texttt{Concat} layers refer to Sigmoid Linear Unit \cite{silu_ramachandran2017searching}, layer normalization~\cite{ba2016layernorm}, and concatenation \mbox{across sequence length, respectively.}
    }
    \label{fig:model_arch}
    \vspace{-0.2cm}
\end{figure}

\modelname{}'s architecture is based on a transformer \cite{attention_is_all_you_need} and is depicted in Figure~\ref{fig:model_arch}.
It draws on the diffusion conditioning paths proposed 
for images in \cite{imagen_saharia2022photorealistic},
and incorporates the relative positional encoding used by wav2vec 2.0 \cite{wav2vec2.0}.
Concretely, the model consists of 24 transformer layers where the vector sequence used for the self-attention block \cite{attention_is_all_you_need} is carefully crafted to incorporate timestep and conditioning information.
For discrete inputs like the character sequence $\mathbf{x}_t$ and the timestep $t$, we first embed them into a continuous vector space using regular learnt embedding layers (for characters) or fixed sinusoidal embeddings \cite{improved_ddpm_2021_sinusoidal_temb} (for the timestep).
The timestep embedding (after a few layers as in Figure~\ref{fig:model_arch}) is summed with each character embedding to \mbox{condition the sequence on the current timestep.}

The frozen WavLM model computes the sequence of features $\mathbf{c}$ associated with the utterance, producing a vector for every 20~ms of the utterance.
To condition \modelname{} on these features $\mathbf{c}$, we adapt the technique proposed for image synthesis in \cite{imagen_saharia2022photorealistic} and compute two streams of information as shown in Figure~\ref{fig:model_arch}.
First, we sum the vector derived from the mean across the entire sequence $\mathbf{c}$ with each character embedding.
And second, we concatenate the entire sequence of vectors $\mathbf{c}$ (after passing it through a few layers) with the sequence of character embeddings, and use this longer sequence as the keys and values for the self-attention block.
Intuitively, the mean-pooled vector is meant to provide \modelname{} with a summary of the entire utterance, 
while providing the full sequence $\mathbf{c}$ to the attention layer allows for the model to learn more fine-grained spelling as it can attend to specific parts of the utterance for each query vector.

This series of operations is all encompassed in a single transformer block, and the full \modelname{} comprises of 24 such blocks and a final softmax projection head to yield the final distribution $p(\blueemph{\hat{\mathbf{x}}_0} | \mathbf{x}_t, \mathbf{c})$.
Finally, since concatenating the entire sequence of WavLM-derived vectors in each transformer block entails a very large memory and compute time cost, we only apply the \texttt{Concat} layer in Figure~\ref{fig:model_arch} in a select few transformer blocks (detailed in Section~\ref{sec:5_exp_setup}).


\section{Diffusion Decoding}\label{sec:4_decoding}

\begin{figure}[t]
    \centering
    \includegraphics[width=0.85\textwidth]{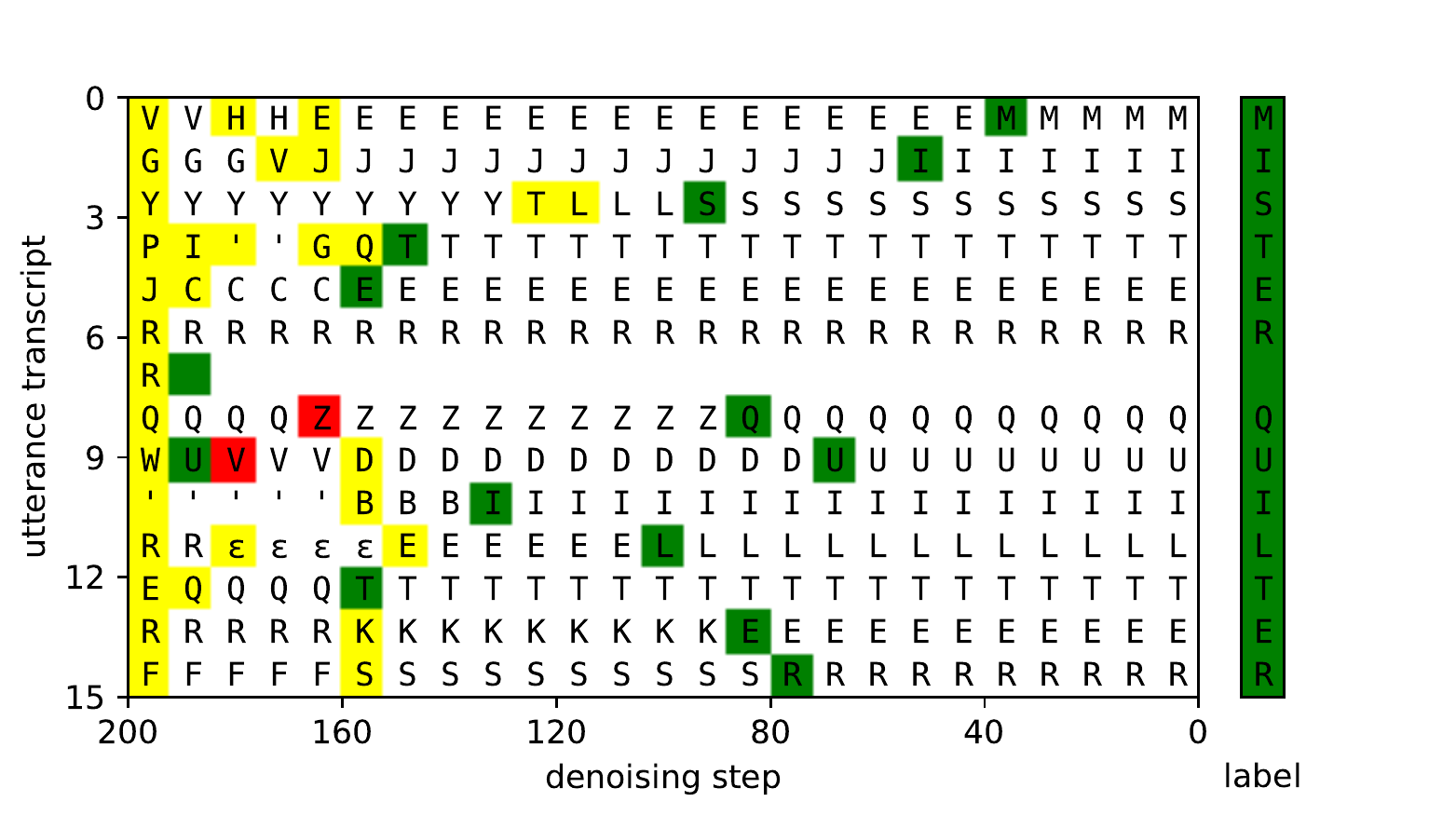}\hspace{-1cm}
    \caption{Example denoising process of the \modelname{} model, starting from a random sequence $\mathbf x_T$ (leftmost column) and denoising until $\mathbf{x}_0$ (rightmost column, excluding label). Green blocks indicate transitions to the ground truth target transcription. Red and yellow blocks indicate transitions from a right character to a wrong character, and a wrong character to another wrong character respectively.}
    \label{fig:denoise_example}
    \vspace{-0.4cm}
\end{figure}

To perform ASR with \modelname{}, we must define the process of decoding a new utterance input through the
diffusion model.
The simplest form of the diffusion decoding process, described briefly in Section~\ref{sec:3_1_conditioning} and shown in Figure~\ref{fig:model_diagram}, iteratively computes $\mathbf x_{t-1}$ given $\mathbf x_{t}$.
An example of this decoding is given in Figure~\ref{fig:denoise_example}, where the speech features $\mathbf c$ contain the linguistic content ``\texttt{MISTER QUILTER}''.
Starting at a random sequence of characters at $t=T=200$,
the model iteratively denoises until the final transcript is reached at $t=0$.
This approach, while effective, often results in errors related to the position of words in the overall sequence.
If the model mistakes word placement early on, it is unable to make corrections because it is unable to easily shift characters.
To this end, we propose new techniques for effectively decoding multinomial diffusion models.

\subsection{Resampling}

RePaint \cite{lugmayr2022repaint} introduced the idea of resampling for the reverse diffusion process to improve inpainting performance during image synthesis.
With RePaint, instead of linearly denoising from $t=T-1$ to $t=0$, they jump back and forth applying both forward and reverse diffusion on a set schedule. 
They found that repeatedly denoising and adding noise (i.e. diffusing) an image improved image generation quality, allowing the model to improve local coherency.
This resampling schedule can be used directly in our decoding.
Concretely, we define two constants, jump length $L$ and number of jumps $J$.
During decoding, we alternate between denoising (applying the reverse function $p(\mathbf{x}_{t-1} | \mathbf{x}_t)$) and diffusing (applying the forward noising process $q(\mathbf{x}_t | \mathbf{x}_{t-1}$)), each lasting $L$ timesteps and repeated $J$ times.
This is repeated until $\frac{T}{L}-1$ times linearly along $t$. 
See \cite{lugmayr2022repaint} for precise details.

\subsection{Sequentially Progressive Diffusion}

Another method introduced in \cite{lugmayr2022repaint} is inpainting or known region conditioning.
The idea of inpainting is to predict the missing pixels under a masked region of an image.
Influenced by this, we considered using a similar approach by decoding word-by-word from the beginning of the transcript.
The predicted words would be then treated as the inpainted region and the model then must predict the rest of the transcript.
While effective, this approach is computationally expensive, as the entire denoising process must be repeated for every word in the transcript.

\begin{figure}[tb]
    \centering
    \includegraphics[width=1\textwidth]{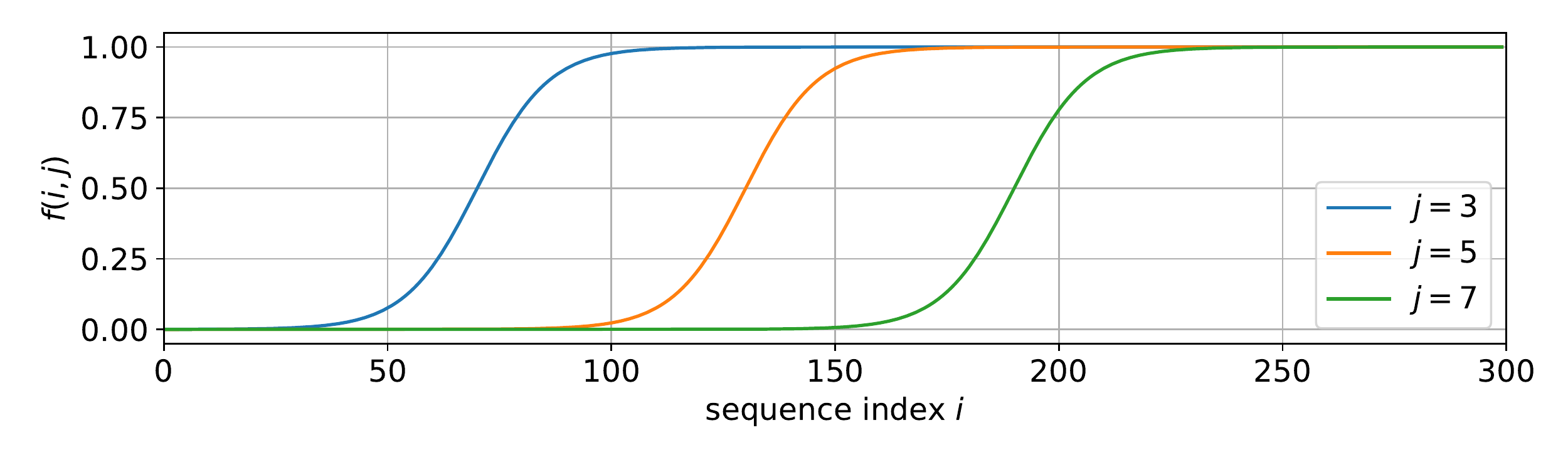}
    \caption{
    Example of sequentially progressive diffusion scaling function $f$ for $J=10$, shown over the character sequence for $j \in \{3,5,7\}$. This function modifies the variance scale based on both resampling jump and sequence index.
    The actual function used is $f(i,j) = \sigma((i-\frac{jN}{J}+2J)/8)$. The constant offset $2J$ is added to ensure the full sequence is diffused at $j=0$, and $8$ was analytically chosen to ensure a smooth overlap between sequential jumps $j$.}
    \label{fig:progressive}
    \vspace{-0.4cm}
\end{figure}

Instead, we develop a new tractable method to implement alongside resampling. We call this sequentially
progressive diffusion:
instead of applying forward diffusion uniformly along the sequence length,
we {now} scale $\beta$ along the length of the transcript.
This allows us to retain earlier parts of the utterance while diffusing{/noising} the later parts.
To scale diffusion noise based on sequence position, we define the scaled diffusion noise schedule $\beta'_{t,i}$ for each character position $i$:
\begin{equation*}
    \beta'_{t,i} = \beta_t \cdot f(i,j)
\end{equation*}
where $f$ defines the scaling function and $j \in \{0,1,\dots,J-1\}$ is the current resample jump.
$f$ is chosen such that the diffusion is applied uniformly at $j=0$, and slides linearly along the sequence length, only diffusing the end of the transcript at $j=J-1$.
We implemented $f$ as a shifted sigmoid as shown in Figure~\ref{fig:progressive}
to make a smooth transition between the retained and diffused regions.

\subsection{Classifier-Free Guidance}

As with other conditional diffusion efforts in the image synthesis domain \cite{imagen_saharia2022photorealistic, dall-e-2_ramesh2022hierarchical}, we utilize classifier-free guidance \cite{clas_free_guidance_ho2021} to improve the alignment between the output transcript and speech features $\mathbf{c}$.
In our context, classifier-free guidance attempts to force \modelname{} to learn both a conditional $\text{\modelname{}}(\mathbf{x}_t, \mathbf{c})$ and unconditional $\text{\modelname{}}(\mathbf{x}_t)$ reverse diffusion process.
This is achieved by randomly dropping out the conditioning information $\mathbf{c}$ with some small probability (in our case 0.1 -- the same as those found to work well in \cite{clas_free_guidance_ho2021,imagen_saharia2022photorealistic}).
In this way, \modelname{} learns to both generate unconditional realistic text and learns to generate text aligned with a transcript.
Then during ASR inference, at each diffusion step, we update the output $p(\hat{\mathbf{x}}_{t-1}\, | \, \mathbf{x}_t, \mathbf{c})$ to move in the direction \textit{from} the unconditional output \textit{to} the conditional output with a `guidance weight' \cite{clas_free_guidance_ho2021}.
More formally, during inference we set:
\begin{equation*}
    p(\hat{\mathbf{x}}_{0}| \mathbf{x}_t, \mathbf{c}) = w\, \text{\modelname{}}(\mathbf{x}_t, \mathbf{c}) + (1-w)\, \text{\modelname{}}(\mathbf{x}_t)
        \label{eq:classifier_guidance}
\end{equation*}
The reasoning is as follows: if the model output (logits of $\hat{\mathbf{x}}_0$) \textit{conditioned} on information about the utterance is $\text{\modelname{}}(\mathbf{x}_t, \mathbf{c})$ and \textit{not conditioned} on the utterance information is $\text{\modelname{}}(\mathbf{x}_t)$, then intuitively the linear direction from $\text{\modelname{}}(\mathbf{x}_t)$ to $\text{\modelname{}}(\mathbf{x}_t, \mathbf{c})$ corresponds to the direction of increasing conditioning information.
We can then improve the strength of the conditioning -- in our case, to improve alignment of output transcript with utterance -- by linearly adjusting the conditional output in this direction \cite{clas_free_guidance_ho2021}.
Note that we apply this linear combination before the output \texttt{softmax} of the model in Figure~\ref{fig:model_arch} to ensure the adjusted output still is a valid probability distribution.
With
$w=1$ in the above equation there is no
guidance while increasing values $w>1$ strengthens the guidance effect.
We found $w=1.5$ to yield the best results 
based on decoding ablations on our validation set (following the same ASR setup as described in Sec.~\ref{sec:5_exp_setup}). We use this setting
\mbox{in all our evaluations.}

\subsection{Full Inference Process}

The full decoding process combines the methods defined above to perform ASR on an utterance from speech features $\mathbf c$.
The core of the inference is resampling with sequentially progressive noise scaling in the forward diffusion steps.
From the same validation decoding ablation experiments, we
found resampling worked best with $J=10$ and $L=10$, which we use for our final resampling decoding in the next section.
The reverse diffusion step utilises classifier-free guidance to improve alignment of the output transcript.
We also note that we can use arbitrary sequence lengths at inference due to the model using relative positional encoding.
In our final inference we use a sequence length $N=400$ to ensure we cover all transcripts in the LibriSpeech test and dev datasets described in Section~\ref{sec:datasets} ($>99\%$ of LibriSpeech transcripts are shorter than 400 characters).
In all our diffusion training and evaluation experiments we set $\beta_t$ according to the cosine noise schedule from \cite{improved_ddpm_2021_sinusoidal_temb} using the recommended value of $s=0.008$.

\section{Experimental Setup}\label{sec:5_exp_setup}

We compare our model against other common ASR models on a standard speech recognition benchmark dataset: LibriSpeech \cite{panayotov2015librispeech}.
Namely, we compare against the high-performing self-supervised speech representations models wav2vec 2.0 \cite{wav2vec2.0}, Conformer \cite{zhang2020pushing_conformer}, and w2v-BERT models \cite{chung2021w2v-bert}.

\subsection{Dataset and Metrics}\label{sec:datasets}
We perform our experiments on the LibriSpeech dataset \cite{panayotov2015librispeech}.
It consists of 960~hours of spoken audiobooks by multiple speakers with varying amounts of noise and audio quality.
We train our model on the full 960~h \texttt{train} split of LibriSpeech and evaluate it on the official \texttt{dev} and \texttt{test}-splits.
For the frozen WavLM model, we use the \texttt{WavLM-Large} pretrained model from the original authors \cite{chen2022wavlm}.
Note that this model has only been pretrained with a masked prediction task on raw audio, and has not been fine-tuned to perform ASR (i.e. it has never been exposed to transcripts of utterances).
To evaluate our model we use the standard ASR metrics of word error rate (WER) and character error rate (CER), however ASR papers typically focus on the WER metric \cite{chen2022wavlm, hubert2021} so we focus on WER for comparison.
We compute the mean WER for our model sampled using various decoding strategies and compare it against the baseline models (described next) using their best results \mbox{reported by the original authors.}

\subsection{Baseline Models}\label{sec:baselines}

We compare against three state-of-the-art models for ASR: wav2vec 2.0 \cite{wav2vec2.0}, Conformer \cite{zhang2020pushing_conformer}, and w2v-BERT \cite{chung2021w2v-bert}.
The first two are large transformers trained in two phases.
First, they are trained in a self-supervised fashion using large amounts of unlabeled audio on a masked token prediction task, and then they are fine-tuned with a CTC objective on the LibriSpeech dataset of labeled (i.e. transcribed) audio.
The w2v-BERT model also follows this pretraining-finetuning setup, but also incorporates additional tricks to improve performance such as self-training with 
noisy student training
\cite{Park2020noisy_student_training}.
These practical techniques to squeeze out additional performance from ASR models have been developed primarily for CTC-type ASR models and are largely undeveloped for diffusion-type ASR models. 
This is because CTC-type models (including all the baseline models) ultimately produce a probability distribution for a character or phoneme being present at a \textbf{certain time} in the utterance.
Meanwhile, our diffusion-type model produces a probability distribution for a character being present at a \textbf{certain position} in the transcript. 
It remains as future work to develop and adapt the practical techniques to improve performance of such diffusion-type models. 

Furthermore, decoding an acoustic model with a typical language model and lexicon has also been developed with these CTC-type models in mind, making them not very effective when applied directly to our model which predicts characters at fixed positions in the transcript.
The primary reason for this is that any insertion or deletion of characters early on in the transcript will cause a substantial change in the predicted likelihood of the rest of the characters in the utterance. 
For example, if the modal model prediction for the first two words is ``\texttt{SQUEZE IT}'',
and the lexicon is naively used to decode the first word to ``\texttt{SQUEEZE }'', then the diffusion acoustic model will have a very high likelihood of ``\texttt{SQUEEZE T}'' and a very low likelihood for the desired correction ``\texttt{SQUEEZE IT}''.

Again, this problem stems from our different way of phrasing the ASR problem as predicting characters at fixed positions in a transcript.
So, if we insert a character early on in the decoding process against the acoustic model's recommendation (i.e. not using its modal prediction), the acoustic likelihood will incentivise the dropping of a character elsewhere in the transcript to retain a high likelihood score for the ground-truth transcript.
Because decent techniques for language model decoding have not yet been developed for diffusion-type models, all experiments in this paper do not use a lexicon or language model---the models that we compare to are also used without a \mbox{language model or lexicon.}

\subsection{\modelname{} Implementation}

\textbf{Layers}:
For the relative positional encoding layer at the input to the transformer, we use the formulation provided in \cite{wav2vec2.0}.
As mentioned in Section~\ref{sec:3_model}, we do not apply the \texttt{Concat} layer in every transformer block.
To save compute resources and to allow for an improved attention weighting (discussed next), we only apply the \texttt{Concat} operation in every 4th block.
So, with 24 transformer blocks, the \texttt{Concat} layer is present in layers $1, 5, 9, 13, 17, 21$.
The self-attention and feed-forward blocks follow the same architecture as in the original attention article \cite{attention_is_all_you_need}.

\textbf{Model hyperparameters}:
Our model uses a 29-sized character alphabet and contains 24 transformer blocks.
The output dimension of all embedding, linear, and attention layers is 768.
The feed-forward blocks have a dimension of $4\times768=3072$ and each self-attention operation uses 8 attention heads.
Transformer attention and feed-forward blocks use a dropout of 0.1, and we also completely dropout all conditioning information with probability 0.1 (in line with classifier-free diffusion guidance \cite{clas_free_guidance_ho2021}).
For the relative positional encoding \cite{wav2vec2.0}, we use a 256-size convolution kernel with 32 convolution groups.
The conditioning sequence $\mathbf{c}$ from \texttt{WavLM-Large} model is defined as the average of the activations of the last 9 layers from the model pretrained on LibriLight \cite{librilight}, since \cite{chen2022wavlm} found these last layers to be most important for representing linguistic information.

\textbf{Optimization}:
We train \modelname{} on the full LibriSpeech 960~h training subset for 350k updates using a batch size of 720 with Adam optimization \cite{adam} with $\beta=(0.9, 0.999)$.
Further, we use a constant learning rate of $3\times10^{-5}$ with a linear warmup of 10k updates and clip the global gradient norm at 10.
As this is an initial foray into ASR with diffusion, we do not use any data augmentation such as SpecAugment. This differs from the baselines, all of which have been trained with substantial data augmentation to further improve performance \cite{wav2vec2.0, zhang2020pushing_conformer, chung2021w2v-bert}.
Even without augmentations, the model's validation performance (WER on internal validation split) was still improving at the end of training -- we hypothesize that training a larger model for longer with all the typical data augmentation techniques used for other CTC-type models will yield further improvements to results listed in the next section.
To demonstrate the level of improvement gained from longer training, we continue training our model up to 462k updates and show its performance compared to the base 350k update model in the next section.
All training is done on three NVIDIA Quadro RTX 6000 devices with mixed FP16/FP32 precision. 

\begin{table}[t!]
    \vspace{-8pt}
    \renewcommand{\arraystretch}{1.2}
    \centering
    \caption{
        WER results on the LibriSpeech dev and test splits for ASR models trained on the full 960~h LibriSpeech training set. Decoding for prior models is done with lexicon-free CTC-decoding \cite{ctc}, while several decoding strategies are applied to our diffusion model (Section~\ref{sec:4_decoding}).
        No language models are used in decoding, and all models are fine-tuned on the full LibriSpeech 960~h training data.
        The unlabeled data used to pretrain each model is specified as the LibriSpeech 960h dataset (LS-960h) \cite{panayotov2015librispeech} or LibriLight 60\,000 hour dataset (LV-60kh) \cite{librilight}.
    }
    \tablecaptionsep
    \footnotesize
    \label{tab:results}
    \begin{tabularx}{0.99\linewidth}{@{}
        L
        c
        @{\hspace{0.3cm}}
        S[table-format=4.0]
        S[table-format=2.1]
        S[table-format=2.1]
        @{\hspace{0.55cm}}
        S[table-format=2.1]
        S[table-format=2.1]@{}}
    \toprule
    &  & & \multicolumn{2}{c}{{dev set \ \ }} & \multicolumn{2}{c}{{test set}} \\
    \cmidrule(lr{0.4cm}){4-5} \cmidrule(l{-0.1cm}){6-7} \vspace{-0.3cm}
    Model & Pretraining & {Params (M)} & {clean} & {other} & {clean} & {other}  \\
    \midrule
    \textit{Including pretraining} \\
    wav2vec 2.0 \texttt{Base}~\cite{wav2vec2.0} & LS-960h & 95 & 3.2 & 8.9 & 3.4 & 8.5 \\
    wav2vec 2.0 \texttt{Large}~\cite{wav2vec2.0} & LS-960h & 317 & 2.6 & 6.5 & 2.8 & 6.3 \\
    wav2vec 2.0 \texttt{Large}~\cite{wav2vec2.0} & LV-60kh & 317 & 2.1 & 4.5 & 2.2 & 4.5 \\
    Conformer XXL \cite{zhang2020pushing_conformer} & LV-60kh & 1000 & 1.6 & 3.2 & 1.6 & 3.3 \\
    w2v-BERT XXL \cite{chung2021w2v-bert} & LV-60kh & 1000 & \ubold 1.5 & \ubold 2.7 & \ubold 1.5 & \ubold 2.8 \\
    \midrule
    \textit{No pretraining} \\
    wav2vec 2.0 \texttt{Large}~\cite{wav2vec2.0} & None & 317 & 2.8 & 7.6 & 3.0 & 8.5 \\
    Conformer L \cite{zhang2020pushing_conformer} & None & 103 & \ubold 1.9 & \ubold 4.4 & \ubold 2.1 & \ubold 4.3 \\
    \midrule
    \textit{TransFusion} (ours, 350k updates) & None & 253 &  &  &  &  \\
     $\; $ basic decoding (Sec.~\ref{sec:3_1_conditioning}) & & & 9.6 & 12.1 & 10.5 & 12.5 \\
     $\;\ +$ classifier free guidance &  &  & 9.4 & 11.7 & 10.2 & 12.2 \\
     $\;\ +$ resampling & & & 8.4 & 10.7 & 9.0 & 11.0 \\ 
     $\;\ +$ sequentially progressive diffusion &  &  & 8.1 & 10.5 & 8.9 & 10.8 \\
     $\;\ +$ further trained to 462k updates & & & \ubold 6.1 & \ubold 8.3 & \ubold 6.7 & \ubold 8.8 \\
    \bottomrule
    \end{tabularx}
    \vspace{-0.2cm}
\end{table}

\section{Results}\label{sec:6_results}

The ASR results are given in Table~\ref{tab:results}.
First we observe that our new model using our best decoding strategy does not beat the current state-of-the-art CTC-type models such as 
Conformer~L
or w2v-BERT.
However, the performance is still considerable given
that this is a first investigation of an entirely new approach to ASR using discrete diffusion.
Concretely, we achieve a \texttt{test-clean} and \texttt{test-other} WER of 6.1\% and 8.8\%, respectively.
The nature of many of our model's mistakes follows 
the issue outlined in Section~\ref{sec:baselines}.
Specifically, often if the model decodes a character by an erroneous insertion/deletion early on in the transcript, it will attempt to drop or insert another erroneous character later on to ensure that the alignment for the rest of the characters is still correct. 
While this 
worsens the WER, the effect on CER is less impactful, where \modelname{}
achieves a CER of 3.2\% and 3.6\% on \texttt{test-clean} and \texttt{test-other}, respectively.

Furthermore,
we observe from Table~\ref{tab:results} that each of our decoding techniques cumulatively improves the results, with our final addition of sequentially progressive diffusion yielding a more than 1.5\% absolute WER improvement over basic decoding.
This demonstrates the effectiveness of our initial decoding methods and suggests that even further improved performance may be achievable given better decoding approaches.
Further training our model to 462k updates also substantially improves performance (last row of Table~\ref{tab:results}), bringing the \texttt{dev-other} subset results above that of wav2vec 2.0 \texttt{Base}.
This suggests that -- in line with our motivation about diffusion scaling in Section~\ref{sec:related} -- even greater performance is likely achievable with increased compute and model sizes.

It is also interesting to observe that the difference in WER for \modelname{} between the less noisy \texttt{dev-}/\texttt{test-clean} sets and the more noisy \texttt{dev-}/\texttt{test-other} sets is much less than all the CTC-type models.
For the CTC models, WER on the clean subsets are often half of the result on the noisy subset, while with the diffusion model the difference is much smaller.
While we are not sure of the precise reason for this, we speculate
that our
decoding method for diffusion-type acoustic models does not draw out the full performance possible from \modelname{}, unlike the powerful CTC decoding possible with CTC-type models.
In other words, if the performance trend of the CTC-type models is representative of the difficulty difference between \texttt{clean} and \texttt{other} subsets, then we should be able to achieve a better clean performance once more optimal decoding methods are developed
for diffusion-type models. 
Finally, while \modelname{} does use features from a pretrained WavLM model, it does not fine-tune the WavLM encoder which provides these features unlike the other methods considered, hence we denote the trained weights of \modelname{} as not \mbox{including any pretraining in Table~\ref{tab:results}.}

\section{Conclusion}

In this paper we proposed \modelname{} -- a 
model that utilizes multinomial diffusion to phrase the task of speech recognition as a conditional discrete diffusion task. 
Our model iteratively denoises an arbitrarily noised transcript until it resembles coherent text corresponding to the transcript of a provided utterance.
This is done by providing speech features associated with the utterance to condition a large transformer model predicting a categorical distribution over a character alphabet.
Since we are the first to phrase ASR in this way, we proposed new methods to decode such diffusion-type ASR models during inference.
We showcase \modelname{}'s performance on the LibriSpeech dataset and compare it to existing state-of-the-art CTC-type models and demonstrate comparable performance.
While we do not outperform the best large CTC-type models we compare to, we achieve a 8.8\%  WER / 3.6\% CER on the LibriSpeech \texttt{test-other} set.
This is noteworthy given the completely new method for ASR proposed here.
Future work will develop better decoding strategies and methods for combining the diffusion acoustic model with a language model.
We will also consider training on standard speech features instead of using WavLM.

\subsubsection*{Acknowledgements} 

All experiments were performed on Stellenbosch University's High Performance Computing (HPC) GPU cluster. This work is supported in part by the National
Research Foundation of \mbox{South Africa (grant no. 120409).}

\bibliographystyle{splncs04}
\bibliography{references}

\end{document}